\pdfoutput=1
\documentclass[sigconf]{acmart}

\newcommand{\hackycomment}[1]{}

 \setcopyright{rightsretained}

\acmConference[FAccT '21]{FAccT '21: The ACM Conference on Fairness, Accountability, and Transparency}{March, 2021}{Toronto, CA (Virtual)}
\acmBooktitle{FAccT '21: ACM Conference on Fairness, Accountability, and Transparency, March 2021, Toronto, CA (Virtual)}

\begin{document}

\title{Measurement and Fairness}

\author{Abigail Z.\ Jacobs}
\email{azjacobs@umich.edu}
\affiliation{
  \institution{University of Michigan}
}

\author{Hanna Wallach}
\email{hanna@dirichlet.net}
\affiliation{
  \institution{Microsoft Research}
}

\copyrightyear{2021}
\acmYear{2021}
\acmConference[FAccT '21]{Conference on Fairness, Accountability, and Transparency}{March 3--10, 2021}{Virtual Event, Canada}
\acmBooktitle{Conference on Fairness, Accountability, and Transparency (FAccT '21), March 3--10, 2021, Virtual Event, Canada}\acmDOI{10.1145/3442188.3445901}
\acmISBN{978-1-4503-8309-7/21/03}

\renewcommand{\shortauthors}{Jacobs and Wallach}

\begin{abstract}
  We propose \emph{measurement modeling} from the quantitative social
  sciences as a framework for understanding fairness in computational
  systems. Computational systems often involve \emph{unobservable
    theoretical constructs}, such as socioeconomic status, teacher
  effectiveness, and risk of recidivism.  Such constructs cannot be
  measured directly and must instead be inferred from measurements of
  observable properties (and other unobservable theoretical
  constructs) thought to be related to them---i.e.,
  \emph{operationalized} via a \emph{measurement model}. This process,
  which necessarily involves making assumptions, introduces the
  potential for mismatches between the theoretical understanding of
  the construct purported to be measured and its
  operationalization. We argue that many of the harms discussed in the
  literature on fairness in computational systems are direct results
  of such mismatches. We show how some of these harms could have been
  anticipated and, in some cases, mitigated if viewed through the lens
  of measurement modeling. To do this, we contribute fairness-oriented
  conceptualizations of \emph{construct reliability} and
  \emph{construct validity} that unite traditions from political
  science, education, and psychology and provide a set of tools for
  making explicit and testing assumptions about constructs and their
  operationalizations. We then turn to fairness itself, an
  \emph{essentially contested construct} that has different
  theoretical understandings in different contexts. We argue that this
  contestedness underlies recent debates about fairness definitions:
  although these debates appear to be about different
  operationalizations, they are, in fact, debates about
  different theoretical understandings of fairness. We show how
  measurement modeling can provide a framework for getting to the core
  of these debates.\looseness=-1
\end{abstract}

\ccsdesc[500]{General and reference~Measurement}
\ccsdesc[500]{General and reference~Validation}
\ccsdesc[500]{General and reference~Reliability}

\keywords{measurement, construct validity, construct reliability, fairness}

\maketitle

\section{Introduction}
\label{sec:intro}

Computational systems have long been touted as having the potential to
counter societal biases and structural inequalities, yet recent work
has demonstrated that they often end up encoding and exacerbating them
instead. Indeed, the literature on fairness in computational systems
has identified many types of fairness-related
harms~\cite{kateNIPSkeynote,barocas2017problem}, as well as many
potential causes, including societal biases reflected in the content
of training datasets, too few data points about particular groups of
people, and assumptions made throughout the development and deployment
lifecycle~\citep[e.g.,][]{olteanu2019social}. However, we argue that
these discussions have overlooked an important subtlety: computational
systems often involve \emph{unobservable theoretical
  constructs}---abstractions that describe phenomena of theoretical
interest, such as socioeconomic status, teacher effectiveness, and
risk of recidivism. Because these constructs cannot be observed, they
cannot be measured directly. Instead, they must instead be inferred
from measurements of observable properties (and other unobservable
theoretical constructs) thought to be related to them---i.e.,
\emph{operationalized} via a \emph{measurement model}. This
process---the \emph{measurement modeling} process--- necessarily
involves making assumptions, thereby introducing the potential for
mismatches between the theoretical understanding of the construct
purported to be measured and its operationalization in a measurement
model. Indeed, we argue that many of the harms studied in the
literature on fairness in computational systems are direct results of
such mismatches.\looseness=-1

Although it is fundamental to the quantitative social sciences,
measurement modeling is largely missing from 1) computer science in
general and 2) the discourse around and literature on fairness in
computational systems in particular.\footnote{We expect that readers
  from the quantitative social sciences will already be intimately
  familiar with measurement modeling. Indeed, some may even wonder why
  we chose to write this paper. However, we emphasize that many
  researchers and practitioners outside of these disciplines have not
  heard of measurement modeling at all, let alone as a framework for
  understanding fairness in computational systems. We therefore view
  this paper as contributing an important bridge between disparate
  disciplines.} As a result, researchers and practitioners are often
inclined to conflate constructs and their operationalizations---i.e.,
to collapse the distinctions between them. But collapsing these
distinctions, either colloquially or epistemically, makes it difficult
to anticipate, let alone mitigate, any possible mismatches. In other
words, collapsing these distinctions elides the space in which
fairness-related harms are most often introduced.

In this paper, we propose measurement modeling as a framework for
understanding fairness in computational systems. Measurement modeling
provides both a language for articulating the distinctions between
constructs and their operationalizations and set of tools---namely
\emph{construct reliability} and \emph{construct validity}---for
surfacing possible mismatches. We argue that these capabilities will
enable researchers and practitioners to 1) better anticipate
fairness-related harms before deploying computational systems in
consequential ways and 2) identify potential causes of
fairness-related harms in ways that reveal concrete, actionable
avenues for mitigating them.\looseness=-1

We further explain how measurement modeling can contribute to recent
debates about fairness definitions. Fairness itself is an unobservable
theoretical construct, albeit one that is \emph{essentially
contested}~\cite{gallie1955essentially,mulligan2016privacy}. In other
words, fairness has multiple context-dependent, and sometimes even
conflicting, theoretical understandings. The contested nature of
fairness makes it inherently hard to measure: If there are multiple
theoretical understandings of a construct, then it is imperative to
articulate which understanding is being operationalized. A failure to
do this makes it difficult to meaningfully compare different
operationalizations because they may be operationalizing different
theoretical understandings. In turn, this makes it difficult to
identify mismatches between fairness as it is theoretically understood
and any given operationalization.\looseness=-1

We argue that although recent debates about fairness definitions have
been framed in terms of operationalizations, they are, in fact,
debates about different theoretical understandings of fairness---i.e.,
about the essentially contested nature of fairness as a construct. We
show how measurement modeling can get to the core of these debates by
providing a language for disentangling debates about different
operationalizations of a single theoretical understanding from debates
about different, and possibly conflicting, theoretical
understandings. This is crucial because debates about different
operationalizations of a single theoretical understanding are debates
about the measurement modeling process---i.e., the assumptions made
when moving from abstractions to mathematics---while debates about
different theoretical understandings are debates about
values.\footnote{We emphasize that this is not to say that values play
  no role in the measurement modeling process---quite the contrary, as
  we explain in the first half of this paper.}  We argue that debates about values
should be explicitly treated as such instead of being couched
indirectly in mathematics, where they are rendered less accessible to
the stakeholders that are most likely to be affected by the
computational systems in question.

In the next section, we give a brief overview of the measurement
modeling process, drawing on well-known examples from the literature
on fairness in computational systems. In
section~\ref{subsec:construct-validity}, we contribute
fairness-oriented conceptualizations of construct reliability and
construct validity, uniting traditions from political science,
education, and psychology. We use the examples introduced in
section~\ref{sec:measurement} to illustrate how these
conceptualizations can be used to surface mismatches between
constructs and their operationalizations. In
section~\ref{sec:fairness}, we turn to fairness itself, addressing
recent debates about fairness definitions. Finally, we conclude with a
discussion in section~\ref{sec:discussion}.\looseness=-1

\vspace{0.5cm}
\section{Making Assumptions}\label{sec:measurement}

\emph{Measurement modeling} plays a central role in the quantitative
social sciences, where many theories involve \emph{unobservable
  theoretical constructs}---i.e., abstractions that describe phenomena
of theoretical interest. For example, researchers in psychology and
education have long been interested in studying intelligence, while
political scientists and sociologists are often concerned with
political ideology and socioeconomic status, respectively. Although
these constructs do not manifest themselves directly in the world, and
therefore cannot be measured directly, they are fundamental to society
and thought to be related to a wide range of observable
properties.\looseness=-1

A \emph{measurement model} is a statistical model that links
unobservable theoretical constructs, \emph{operationalized} as latent
variables, and observable properties---i.e., data about the
world~\cite{jackman2008oxford}. In this section, we give a brief
overview of the measurement modeling process, starting with two
comparatively simple examples---measuring height and measuring
socioeconomic status---before moving on to three well-known examples
from the literature on fairness in computational systems. We emphasize
that our goal in this section is not to provide comprehensive
mathematical details for each of our five examples, but instead to
introduce key terminology and, more importantly, to highlight that the
measurement modeling process necessarily involves making assumptions
that must be made explicit and tested before the resulting
measurements are used.

\subsection{Measuring Height}

We start by formalizing the process of measuring the height of a
person---a property that is typically thought of as being observable
and therefore easy to measure directly. There are many standard tools
for measuring height, including rulers, tape measures, and height
rods. Indeed, measurements of observable properties like height are
sometimes called \emph{representational measurements} because they are
derived by ``representing physical objects [such as people and rulers]
and their relationships by
numbers''~\cite{hand2004measurement}. Although the height of a person
is not an unobservable theoretical construct, for the purpose of
exposition, we refer to the abstraction of height as a~construct
$\mathcal{H}$ and then operationalize $\mathcal{H}$ as a latent
variable $h$.

Despite the conceptual simplicity of height---usually understood to be
the length from the bottom of a person's feet to the top of their head
when standing erect---measuring it involves making several
assumptions, all of which are more or less appropriate in different
contexts and can even affect different people in different ways. For
example, should a person's hair contribute to their height? What about
their shoes? Neither are typically viewed as being an intrinsic part
of a person's height, yet both contribute to a person's effective
height, which may matter more in ergonomic contexts. Similarly, if a
person uses a wheelchair, then their standing height may be less
relevant than their sitting height. These assumptions must be made
explicit and tested before using any measurements that depend upon
them.\looseness=-1

In practice, it is not possible to obtain error-free measurements of a
person's height, even when using standard tools. For
example, when using a ruler, the angle of the ruler, the granularity
of the marks, and human error can all result in erroneous
measurements. However, if we take $N$ measurements
$\{\hat{h}_n\}_{n=1}^N$ of a person's height, then provided that the
ruler is not statistically biased, the average will converge to the
person's ``true'' height $h$ with probability one as
$N\rightarrow\infty$. Specifically, we say that the person's true
height---the latent variable $h$---influences the measurements
$\{\hat{h}_n\}_{n=1}^N$---a set of $N$ observed variables. We can
represent this relationship using a graphical model, as shown
below. Observed variables are shaded, latent variables are unshaded,
and annotated boxes denote replication.\looseness=-1

\vspace{0.15cm}
\begin{centering}
  \includegraphics[width=0.2\linewidth]{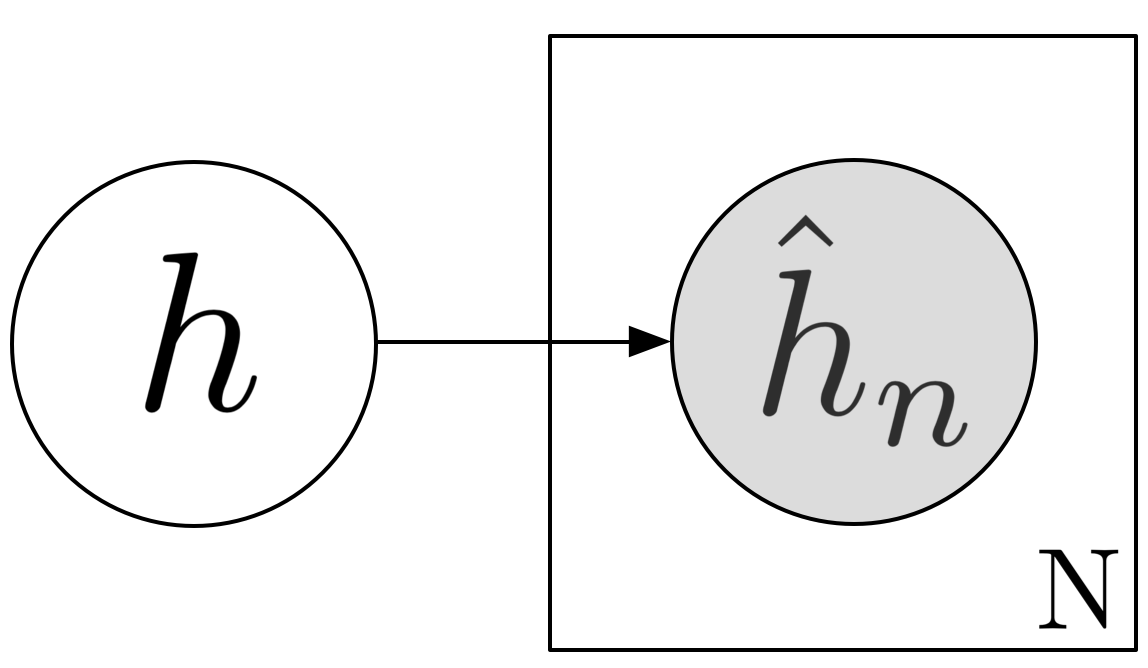}\\
\end{centering}
 \vspace{0.1cm}

Alternatively, we can represent the $n^{\textrm{th}}$ measurement as
\begin{equation}\label{eqn:height-simple}
  \hat{h}_n = h + \epsilon_n,
\end{equation}
where $\epsilon_n$ is the error associated with $\hat{h}_n$. In many
contexts, it is reasonable to assume that the errors associated with
$\{\hat{h}_n\}_{n=1}^N$ are well-behaved---i.e., normally distributed,
statistically unbiased, and possessing small variance
$\sigma^2$. Under this assumption, $\epsilon_n \sim
\mathcal{N}(0,\sigma^2)$---i.e., each error is drawn from a zero-mean
normal distribution with variance $\sigma^2$. This means that
$\frac{1}{N}\sum_{n=1}^N \hat{h}_n \rightarrow h$ with probability one
as $N \rightarrow \infty$---a property of a consistent
estimator. Equation~\ref{eqn:height-simple} is equivalent to the
graphical model representation above provided that $\hat{h}_n \sim
\mathcal{N}(h, \sigma^2)$ for $n=1, \ldots, N$. Borrowing from the
economics literature, we refer to models that formalize the
relationships between~measurements and their errors as \textit{measurement
  error models}.\looseness=-1

We emphasize that even specifying a measurement error model involves making
assumptions. For instance, in some contexts, the errors associated
with $\{\hat{h}_n\}_{n=1}^N$ may not be not well-behaved and may even
be correlated with demographic factors, such as race or gender. As an example, suppose that $\{\hat{h}_n\}_{n=1}^N$ come not from a ruler
but instead from self-reports on dating websites. It might initially
seem reasonable to assume that the corresponding errors are well-behaved in this
context. However, Toma et al.~\cite{toma2008separating} found that
although men and women both over-report their height on dating
websites, men are more likely to over-report and to over-report by a
larger amount. Toma et al.\ suggest this is strategic, likely
representing intentional deception. However, regardless of the cause, these
errors are not well-behaved and are correlated with gender. Assuming
that they are well-behaved will yield inaccurate
measurements.\looseness=-1

\subsection{Measuring Socioeconomic Status}
\label{sec:income_as_proxy}

We now consider the process of measuring a person's socioeconomic
status (SES). From a theoretical perspective, a person's SES is
understood as encompassing their social and economic position in
relation to others. Unlike a person's height, their SES is unobservable, so it cannot
be measured directly and must instead be inferred from measurements of
observable properties (and other unobservable theoretical constructs)
thought to be related to it, such as income, wealth, education, and
occupation. Measurements of phenomena like SES are sometimes called
\emph{pragmatic measurements} because they are designed to capture
particular aspects of a phenomenon for particular
purposes~\cite{hand2004measurement}. We refer to the abstraction of
SES as a~construct $\mathcal{S}$ and then operationalize
$\mathcal{S}$ as a latent variable $s$.\looseness=-1

The simplest way to measure a person's SES is to use an observable
property---like their income---as a proxy for it. Letting the
construct $\mathcal{I}$ represent the abstraction of income and
operationalizing $\mathcal{I}$ as a latent variable $i$, this means
specifying a both measurement model that links $s$ and $i$ and a
measurement error model. For example, if we assume that $s$ and $i$
are linked via the identity function---i.e., that $s = i$---and we
assume that it is possible to obtain error-free measurements of a
person's income---i.e., that $\hat{i} = i$---then $s =
\hat{i}$.\looseness=-1

Like the previous example, this example highlights that the
measurement modeling process necessarily involves making
assumptions. Indeed, there are many other measurement models that use
income as a proxy for SES but make different assumptions about the
specific relationship between them, including other deterministic
models---e.g., $i = 4\times s^2$---and random models---e.g., $i \sim
N(s, \sigma^2)$.

Similarly, there are many other measurement error models that make
different assumptions about the errors that occur when measuring a
person's income. For example, if we measure a person's monthly income
by totaling the wages deposited into their account over a single
one-month period, then we must use a measurement error model that
accounts for the possibility that the timing of the one-month period
and the timings of their wage deposits may not be aligned. Using a
measurement error model that does not account for this
possibility---e.g., using $\hat{i} = i$---will yield inaccurate
measurements. Human Rights Watch reported exactly this scenario in the
context of the Universal Credit benefits system in the
U.K.~\cite{hrw2020uk}: The system measured a claimant's monthly income
using a one-month rolling period that began immediately after they
submitted their claim without accounting for the possibility described
above. This meant that the system ``might detect that an individual
received a \textsterling 1000 paycheck on March 30 and another
\textsterling 1000 on April 29, but not that each \textsterling 1000
salary is a monthly wage [leading it] to compute the individual's
benefit in May based on the incorrect assumption that their combined
earnings for March and April (i.e., \textsterling 2000) are~their~monthly wage,'' denying them much-needed resources.\looseness=-1

Moving beyond income as a proxy for SES, there are arbitrarily many
ways to operationalize SES via a measurement model, incorporating both
measurements of observable properties, such as wealth, education, and
occupation, as well as measurements of other unobservable theoretical
constructs, such as cultural capital.\looseness=-1

\subsection{Measuring Teacher Effectiveness}
\label{subsec:EVAAS}

We turn next to three well-known examples from the literature on
fairness in computational systems, starting with the value-added
models that are used by many school districts to measure the ``value
added'' to students' academic progress by a teacher---i.e., teacher
effectiveness. At the risk of stating the obvious, teacher
effectiveness is an unobservable theoretical construct that cannot be
measured directly and must instead be inferred from measurements of
observable properties (and other unobservable theoretical
constructs). Many organizations have developed models that purport to
measure teacher effectiveness. For instance, SAS's Education
Value-Added Assessment System (EVAAS), which is widely used across the
U.S., implements two models---a multivariate response model (MRM)
intended to be used when standardized tests are given to students in
consecutive grades and a univariate response model intended to be used
in other testing contexts. Although the models differ in terms of
their mathematical details, both use changes in students'
test scores (an observable property) as a proxy for teacher
effectiveness.\looseness=-1

We focus on the EVAAS MRM in this example, though we emphasize that
many of the assumptions that it makes---most notably that students'
test scores are a reasonable proxy for teacher effectiveness---are
common to other value-added models. When describing the MRM, the EVAAS
documentation states that ``each teacher is assumed to
be the state or district average in a specific year, subject, and
grade until the weight of evidence pulls him or her above or below
that average''~\cite{evaasDoc}. The MRM operationalizes the
effectiveness of teacher $t$ for subject $j$, grade $k$, and year\footnote{Year $l$ is only present for accounting purposes; it is
  effectively determined by $(i,j,k)$.} $l$ as~a~latent variable $q_{tjkl}$ via the following measurement model:\looseness=-1
\begin{align}
  q_{tjkl} &= \mu_{jkl} + \sum_{i} \tau_{ijklt}\\
  y_{ijkl} &= \mu_{jkl} + \left(\sum_{k^* <=k}
  \sum_{t=1}^{T_{ijk^*l^*}} w_{ijk^*l^*t} \tau_{ijk^*l^*t} \right),
\end{align}
where $\mu_{jkl}$ is the state or district's estimated mean score for
subject $j$ in grade $k$ in year $l$, $\tau_{ijklt} \sim
\mathcal{N}(0, \sigma^2_{jkl})$ is the effect of teacher $t$ on
student $i$ for subject $j$ in grade $k$ in year $l$, $y_{ijkl}$ is
student $i$'s test score for subject $j$ in grade $k$ in year $l$,
$w_{ijk^*l^*t}$ is the fraction of student $i$'s instructional time
attributed to teacher $t$ for subject $j$ in grade $k$ in year $l$,
and $\tau_{ijk^*l^*t}$ is the effect of teacher $t$ on student $i$ for
subject $j$ in grade $k^*$ in year $l^*$. The teacher effects are
assumed to be normally distributed around zero, with a block diagonal
covariance matrix that contains a block for each subject $j$, grade $k$, and
year $l$. The MRM's measurement error model is $\hat{y}_{ijkl} = y_{ijkl}
+ \epsilon_{ijkl}$, where $\epsilon_{ijkl}$ is assumed to be normally
distributed around zero.

{\looseness=-1 \spaceskip= 2pt plus 1pt minus 1.5pt
As well as assuming that teacher effectiveness is fully captured by
students' test scores, this model makes several other assumptions,
which we make explicit here for expository purposes: 1) that student
$i$'s test score for subject $j$ in grade $k$ in year $l$ is a
function of only their current and previous teachers' effects; 2) that
the effectiveness of~teacher $t$ for subject $j$, grade $k$, and year
$l$ depends on their effects on all of their students; 3) that student
$i$'s instructional time for subject~$j$ in grade $k$ in year $l$ may
be shared between teachers; and 4) that a~teacher may be effective in
one subject but ineffective in another.\looseness=-1}

\subsection{Measuring Risk of Recidivism}\label{subsec:recidivism}

We now consider another well-known example from the literature on
fairness in computational systems: the risk assessment models used in
the U.S. justice system to measure a defendant's risk of
recidivism. There are many such models, but we focus here on
Northpointe's Correctional Offender Management Profiling for
Alternative Sanctions (COMPAS), which was the subject of an
investigation by Angwin et al.~\cite{angwin2016machine} and many
academic papers
\citep[e.g.,][]{corbett2016computer,berk2017fairness,kleinberg2016inherent}.\looseness=-1

COMPAS draws on several criminological theories to operationalize a
defendant's risk of recidivism using measurements of a variety of
observable properties (and other unobservable theoretical constructs)
derived from official records and interviews. These properties and
measurements span four different dimensions: prior criminal history,
criminal associates, drug involvement, and early indicators of
juvenile delinquency problems~\cite{recentCOMPASguide}. The
measurements are combined in a regression model, which outputs a score
that is converted to a number between one and ten with ten being the
highest risk. Although the full mathematical details of COMPAS are
not readily available, the COMPAS documentation mentions numerous
assumptions, the most important of which is that recidivism is defined
as ``a new misdemeanor or felony arrest within two years.'' We discuss
the implications of this assumption in
sections~\ref{subsec:construct-validity}
and~\ref{sec:fairness}.\looseness=-1

\subsection{Measuring Patient Benefit}\label{subsec:patient-benefit}

Finally, we turn to a different type of risk assessment model, used in
the U.S. healthcare system to identify the patients that will benefit
the most from enrollment in high-risk care management programs---i.e.,
programs that provide access to additional resources for patients with
complex health issues. As explained by Obermeyer et al., these models
assume that ``those with the greatest care needs will benefit the most
from the programs''~\cite{obermeyer2019health}.  Furthermore, many of
them~operationalize greatest care needs as greatest care costs.

This assumption---i.e., that care costs are a reasonable proxy for
care needs---transforms the difficult task of measuring the extent to
which a patient will benefit from a program (an unobservable
theoretical construct) into the simpler task of predicting their
future care costs based on their past care costs (an observable
property). However, this assumption masks an important confounding
factor: patients with comparable past care needs but different access
to care will likely have different past care costs. As we explain in
the next section, even without considering any other details of these
models, this assumption can lead to fairness-related
harms~\cite{obermeyer2019health}.

\section{Testing Assumptions}
\label{subsec:construct-validity}

As we explained in the previous section, the measurement modeling
process necessarily involves making assumptions. However, these
assumptions must be made explicit and tested before the resulting
measurements are used. Leaving them implicit or untested obscures any
possible mismatches between the theoretical understanding of the
construct purported to be measured and its operationalization, in turn
obscuring any resulting fairness-related harms. In this section, we
show how measurement modeling provides a set of tools for surfacing
such mismatches. These tools differentiate measurement modeling from
statistical modeling and machine learning as they are typically used
in computer science. We therefore argue that it is these tools that
give measurement modeling its power as a framework for understanding
fairness in computational systems.

In contrast to computer scientists, who focus primarily on
out-of-sample prediction, quantitative social scientists typically
test their assumptions by assessing \emph{construct reliability} and
\emph{construct validity}. Quinn et al.\ describe these concepts as
follows: ``The evaluation of any measurement is generally based on its
reliability (can it be repeated?) and validity (is it
right?). Embedded within the complex notion of validity are
interpretation (what does it mean?) and application (does it
`work?')''~\cite{quinn2010analyze}.  We contribute fairness-oriented
conceptualizations of construct reliability and construct validity
that unite traditions from political science, education, and
psychology by drawing on the work of Quinn et
al.~\cite{quinn2010analyze}, Jackman~\cite{jackman2008oxford},
Messick~\cite{messick1987validity}, and
Loevinger~\cite{loevinger1957objective}, among others. We illustrate
these conceptualizations using the five examples introduced in the
previous section, arguing that they constitute a set of tools that
will enable researchers and practitioners to 1) better anticipate
fairness-related harms that can be obscured by focusing primarily on
out-of-sample prediction, and 2) identify potential causes of
fairness-related harms in ways~that reveal concrete, actionable
avenues for mitigating them.\looseness=-1

\subsection{Construct Reliability}\label{subsec:reliability}

We start by describing construct reliability---a concept that is
roughly analogous to the concept of precision (i.e., the inverse of
variance) in statistics~\cite{jackman2008oxford}. Assessing construct
reliability means answering the following question:
do similar inputs
to a measurement model, possibly presented at different points in
time, yield similar outputs? If the answer to this question is no,
then the model lacks reliability, meaning that we may not want to use
its measurements. We note that a lack of reliability can also make it challenging to assess construct validity. Although different
disciplines emphasize different aspects of construct reliability, we
argue that there is one aspect---namely \emph{test--retest
  reliability}, which we describe below---that is especially~relevant
in the context of fairness in computational
systems.\footnote{\emph{Inter-rater reliability} refers to the extent
  to which multiple raters, experts, judges, or annotators agree in
  their assessments (i.e., outputs) when given the same task (i.e.,
  inputs). Although inter-rater reliability plays a key role when
  measuring unobservable theoretical constructs via qualitative
  methods, it is less immediately relevant when using
  quantitative methods. For this reason, we omit inter-rater
  reliability from our conceptualization of construct reliability. We
  also omit \emph{inter-item reliability}, which refers to the extent
  to which the inputs to a measurement model are correlated with one
  another. Although we believe that inter-item reliability may be
  relevant in the context of fairness in computational systems, we
  could not find any examples of fairness-related~harms that would
  likely have been surfaced by assessing inter-item
  reliability.\looseness=-1}\looseness=-1

\subsubsection{Test--retest reliability}
\label{sec:test--retest}

Test--retest reliability refers to the extent to which measurements of
an unobservable theoretical construct, obtained from a measurement
model at different points in time, remain the same, assuming that the
construct has not changed.\looseness=-1

For example, when measuring a person's height, operationalized as the
length from the bottom of their feet to the top of their head when
standing erect, measurements that vary by several inches from one day
to the next would suggest a lack of test--retest
reliability. Investigating this variability might reveal its cause
to be the assumption that a person's shoes should contribute to their
height.\looseness=-1

As another example, many value-added models, including the EVAAS MRM,
have been criticized for their lack of test--retest reliability.  For
instance, in \emph{Weapons of Math
  Destruction}~\cite{oneil2016weapons}, O'Neil described how
value-added models often produce measurements of teacher effectiveness
that vary dramatically between years. In one case, she described Tim
Clifford, an accomplished and respected New York City middle school
teacher with over 26 years of teaching experience. For two years in a
row, Clifford was evaluated using a value-added model, receiving a
score of 6 out of 100 in the first year, followed by a score of 96 in
the second. It is extremely unlikely that teacher effectiveness would
vary so dramatically from one year to the next. Instead, this
variability, which suggests a lack of test--retest reliability, points
to a possible mismatch between the construct purported to be measured
and its operationalization.

As a third example, had the developers of the Universal Credit
benefits system described in section~\ref{sec:income_as_proxy}
assessed the test--retest reliability of their system by checking that
the system's measurements of a claimant's income were the same no
matter when their one-month rolling period began, they might have
anticipated (and even mitigated) the harms revealed by Human Rights
Watch~\cite{hrw2020uk}.

Finally, we note that an apparent lack of test--retest reliability
does not always point to a mismatch between the theoretical
understanding of the construct purported to be measured and its
operationalization. In some cases, an apparent lack of test--retest
reliability can instead be the result of unexpected changes to the
construct itself. For example, although we typically think of a
person's height as being something that remains relatively static over
the course of their adult life, most people actually get shorter as
they get older.\looseness=-1

\subsection{Construct Validity}\label{subsec:types-of-validity}

Whereas construct reliability is roughly analogous to the concept of
precision in statistics, construct validity is roughly analogous to
the concept of statistical
unbiasedness~\cite{jackman2008oxford}. Establishing construct validity
means demonstrating, in a variety of ways, that the measurements
obtained from measurement model are both meaningful and useful: Does
the operationalization capture all relevant aspects of the construct
purported to be measured? Do the measurements look plausible? Do they
correlate with other measurements of the same construct? Or do they
vary in ways that suggest that the operationalization may be
inadvertently capturing aspects of other constructs? Are the
measurements predictive of measurements of any relevant observable
properties (and other unobservable theoretical constructs) thought to
be related to the construct, but not incorporated into the
operationalization?  Do the measurements support known hypotheses
about the construct? What are the consequences of using the
measurements---including any societal
impacts~\cite{selbst2018fairness,messick1987validity}. We emphasize
that a key feature, not a bug, of construct validity is that it is not
a yes/no box to be checked: construct validity is always~a matter of
degree, to be supported by critical
reasoning~\cite{loevinger1957objective}. \looseness=-1

Different disciplines have different conceptualizations of construct
validity, each with its own rich history. For example, in some
disciplines, construct validity is considered distinct from
\emph{content validity} and \emph{criterion validity}, while in other
disciplines, content validity and criterion validity are grouped under
the umbrella of construct validity. Our conceptualization unites
traditions from political science, education, and psychology by
bringing together the seven different aspects of construct validity
that we describe below. We argue that each of these aspects plays a
unique and important~role in understanding fairness in computational
systems.

\subsubsection{Face validity}

Face validity refers to the extent to which the measurements obtained
from a measurement model look plausible---a ``sniff test'' of sorts.
This aspect of construct validity is inherently subjective, so it is
often viewed with skepticism if it is not supplemented with other,
less subjective evidence. However, face validity is a prerequisite for
establishing construct validity: if the measurements obtained from a
measurement model aren't facially valid, then they are unlikely to
possess other aspects of construct validity.\looseness=-1

It is likely that the models described in
section~\ref{sec:measurement} would yield measurements that are, for
the most part, facially valid. For example, measurements obtained by
using income as a proxy for SES would most likely possess face
validity. SES and income are certainly related and, in general, a
person at the high end of the income distribution (e.g., a CEO) will
have a different SES than a person at the low end (e.g., a
barista). Similarly, given that COMPAS draws on several criminological
theories to operationalize a defendant's risk of recidivism, it is
likely that the resulting scores would look plausible.\looseness=-1

One exception to this pattern is the EVAAS MRM.  Some scores may look
plausible---after all, students' test scores are not unrelated to
teacher effectiveness---but the dramatic variability that we
described above in the context of test--retest reliability is
implausible.\looseness=-1

\subsubsection{Content validity}\label{subsubsec:content-validity}

Content validity refers to the extent to which an operationalization
wholly and fully captures the substantive nature of the construct
purported to be measured. This aspect of construct validity has three
sub-aspects, which we describe below.\looseness=-1

The first sub-aspect relates to the construct's \emph{contestedness}.
If a
construct is essentially contested then it has multiple
context-dependent, and sometimes even conflicting, theoretical
understandings. Contestedness makes it inherently hard to assess
content validity: if a construct has multiple theoretical
understandings, then it is unlikely that a single operationalization
can wholly and fully capture its substantive nature in a meaningful
fashion. For this reason, some traditions make a single theoretical
understanding of the construct purported to be measured a prerequisite
for establishing content
validity~\cite{jackman2008oxford,hand2004measurement}. However, other
traditions simply require an articulation of which understanding is
being operationalized~\cite{sireci1998construct}. We take the
perspective that the latter approach is more practical because it is
often the case that unobservable theoretical constructs are
essentially contested, yet we still wish to measure them. Indeed,
fairness itself is one such construct, as we explain in
section~\ref{sec:fairness}.\looseness=-1

Of the models described in section~\ref{sec:measurement}, most are
intended to measure unobservable theoretical constructs that are
(relatively) uncontested.  One possible exception is patient benefit,
which can be understood in a variety of different ways.  However, the
understanding that is operationalized in most high-risk care
management enrollment models is clearly articulated. As Obermeyer et
al.~explain, ``[the patients] with the greatest care needs will
benefit the most'' from enrollment in high-risk care management
programs~\cite{obermeyer2019health}.\looseness=-1

The second sub-aspect of content validity is sometimes known as
\emph{substantive validity}. This sub-aspect moves beyond the
theoretical understanding of the construct purported to be measured
and focuses on the measurement modeling process---i.e., the
assumptions made when moving from abstractions to
mathematics. Establishing substantive validity means demonstrating
that the operationalization incorporates measurements of those---and
only those---observable properties (and other unobservable theoretical
constructs, if appropriate) thought to be related to the construct.

For example, although a person's income contributes to their
SES, their income is by no means the only contributing factor. Wealth,
education, and occupation all affect a person's SES, as do other
unobservable theoretical constructs, such as cultural capital. For
instance, an artist with significant wealth but a low income should
have a higher SES than would be suggested by their income
alone.\looseness=-1

As another example, COMPAS defines recidivism as ``a new misdemeanor
or felony arrest within two years.'' By assuming that arrests are a
reasonable proxy for crimes committed, COMPAS fails to account for
false arrests or crimes that do not result in
arrests~\cite{roberts2018arrests}. Indeed, no computational system can
ever wholly and fully capture the substantive nature of crime by using
arrest data as a proxy.\looseness=-1

Similarly, high-risk care management enrollment models assume that care costs are a reasonable proxy for care needs.
However, a patient's care needs reflect their underlying health status, while their care costs reflect both their access to care and their health status.\looseness=-1

Finally, establishing \emph{structural validity}, the third sub-aspect
of content validity, means demonstrating that the operationalization
captures the structure of the relationships between the incorporated
observable properties (and other unobservable theoretical constructs,
if appropriate) and the construct purported to be measured, as well as
the interrelationships between
them~\cite{loevinger1957objective,messick1987validity}.\looseness=-1

In addition to assuming that teacher effectiveness is wholly and fully
captured by students' test scores---a clear threat to substantive
validity~\cite{amrein2008methodological}---the EVAAS MRM assumes that
a student's test score for subject $j$ in grade $k$ in year $l$ is
approximately equal to the sum of the state or district's estimated
mean score for subject $j$ in grade $k$ in year $l$ and the student's
current and previous teachers' effects (weighted by the fraction of
the student's instructional time attributed to each teacher). However,
this assumption ignores the fact that, for many students, the
relationship may be more complex.\looseness=-1

\subsubsection{Convergent validity}
\label{sec:convergent}

Convergent validity refers to the extent to which the measurements
obtained from a measurement model correlate with other measurements of
the same construct, obtained from measurement models for which
construct validity has already been established. This aspect of
construct validity is typically assessed using quantitative methods,
though doing so can reveal qualitative differences between different
operationalizations.

We note that assessing convergent validity raises an inherent
challenge: ``If a new measure of some construct differs from an
established measure, it is generally viewed with skepticism. If a new
measure captures exactly what the previous one did, then it is
probably unnecessary''~\cite{quinn2010analyze}. The measurements
obtained from a new measurement model should therefore deviate only
slightly from existing measurements of the same construct. Moreover,
for the model to be viewed as possessing convergent validity, these
deviations must be well justified and supported by critical
reasoning.\looseness=-1

Many value-added models, including the EVAAS MRM, lack convergent
validity~\cite{amrein2008methodological}. For example, in
\emph{Weapons of Math Destruction}~\cite{oneil2016weapons}, O'Neil
described Sarah Wysocki, a fifth-grade teacher who received a low
score from a value-added model despite excellent reviews from her
principal, her colleagues, and her students' parents.\looseness=-1

As another example, measurements of SES obtained from the model
described in section~\ref{sec:income_as_proxy} and measurements of SES
obtained from the National Committee on Vital and Health Statistics
would likely correlate somewhat because both operationalizations
incorporate income. However, the latter operationalization also
incorporates measurements of other observable properties, including
wealth, education, occupation, economic pressure, geographic location,
and family size~\cite{ncvhs2012}. As a result, it is also likely that
there would also be significant differences between the two sets of
measurements. Investigating these differences might reveal aspects of
the substantive nature of SES, such as wealth or education, that are
missing from the model described in
section~\ref{sec:income_as_proxy}. In other words, and as we described
above, assessing convergent validity can reveal qualitative
differences between different operationalizations of a
construct.\looseness=-1

We emphasize that assessing the convergent validity of a measurement
model using measurements obtained from measurement models that have
not been sufficiently well validated can yield a false sense of
security. For example, scores obtained from COMPAS would likely
correlate with scores obtained from other models that similarly use
arrests as a proxy for crimes committed, thereby obscuring the threat
to content validity that we described above.\looseness=-1

\subsubsection{Discriminant validity}

Discriminant validity refers to the extent to which the measurements
obtained from a measurement model vary in ways that suggest that the
operationalization may be inadvertently capturing aspects of other
constructs. Measurements of one construct should only correlate with
measurements of another to the extent that those constructs are
themselves related. As a special case, if two constructs are totally
unrelated, then there~should be no correlation between their
measurements~\cite{hand2004measurement}.

Establishing discriminant validity can be especially challenging when
a construct has relationships with many other constructs. SES, for
example, is related to almost all social and economic constructs,
albeit to varying extents. For instance, SES and gender are somewhat
related due to labor segregation and the persistent gender wage gap,
while SES and race are much more closely related due to historical
racial inequalities resulting from structural racism. When assessing
the discriminant validity of the model described in
section~\ref{sec:income_as_proxy}, we would therefore hope to find
correlations that reflect these relationships. If, however, we instead
found that the resulting measurements were perfectly correlated with
gender or uncorrelated with race, this would suggest a lack of
discriminant validity.\looseness=-1

As another example, Obermeyer et al.\ found a strong correlation
between measurements of patients' future care needs, operationalized
as future care costs, and
race~\cite{obermeyer2019health}. According to their analysis of one
model, only 18\% of the patients identified for enrollment in
high-risk care management programs were Black. This correlation
contradicts expectations. Indeed, given the enormous racial health
disparities in the U.S., we might even expect to see the opposite
pattern. Further investigation by Obermeyer et al.\ revealed that this
threat to discriminant validity was caused by the confounding factor
that we described in section~\ref{subsec:patient-benefit}: Black and
white patients with comparable past care needs had radically different
past care costs---a consequence of structural racism that was then
exacerbated by the model.\looseness=-1

\subsubsection{Predictive validity}

Predictive validity refers to the extent to which the measurements
obtained from a measurement model are predictive of measurements of
any relevant observable properties (and other unobservable theoretical
constructs) thought to be related to the construct purported to be
measured, but not incorporated into the operationalization. Assessing
predictive validity is therefore distinct from out-of-sample
prediction~\cite{grimmer2012comment,mullainathan2017machine}. Predictive validity
can be assessed using either qualitative or quantitative methods. We note
that in contrast to the aspects of construct validity that we
discussed above, predictive validity is primarily concerned~with the
utility of the measurements, not their meaning.

As a simple illustration of predictive validity, taller people
generally weigh more than shorter people. Measurements of a person's
height should therefore be somewhat predictive of their
weight.\looseness=-1

Similarly, a person's SES is related to many observable
properties---ranging from purchasing behavior to media
appearances---that are not always incorporated into models for
measuring SES. Measurements obtained by using income as a proxy for
SES would most likely be somewhat predictive of many of these
properties, at least for people at the high and low ends of the income
distribution.\looseness=-1

We note that the relevant observable properties (and other
unobservable theoretical constructs) need not be ``downstream'' of
(i.e., thought to be influenced by) the construct. Predictive validity
can also be assessed using ``upstream'' properties and constructs,
provided that they are not incorporated into the
operationalization. For example, Obermeyer et al.\ investigated the
extent to which measurements of patients' future care needs, operationalized as
future care costs, were predictive of patients' health statuses (which
were not part of the model that they
analyzed)~\cite{obermeyer2019health}. They found that Black and white
patients with comparable future care costs did not have comparable
health statuses---a threat to predictive validity caused (again) by
the confounding factor described in
section~\ref{subsec:patient-benefit}.

\subsubsection{Hypothesis validity}

Hypothesis validity refers to the extent to which the measurements
obtained from a measurement model support substantively interesting
hypotheses about the construct purported to be measured. Much like
predictive validity, hypothesis validity is primarily concerned with
the utility of the measurements. We note that the main distinction
between predictive validity and hypothesis validity hinges on the
definition of ``substantively interesting hypotheses.'' As a result,
the distinction is not always clear cut. For example, is the
hypothesis ``People with higher SES are more likely to be mentioned in
the New York Times'' sufficiently substantively interesting? Or would
it be more appropriate to use the hypothesized relationship to assess
predictive validity? For this reason, some traditions merge predictive
and hypothesis
validity~\citep[e.g.,][]{jackman2008oxford}.\looseness=-1

Turning again to the value-added models discussed in
section~\ref{subsec:EVAAS}, it is extremely unlikely that the
dramatically variable scores obtained from such models would support
most substantively interesting hypotheses involving teacher
effectiveness, again suggesting a possible mismatch between the
theoretical understanding of the construct purported to be measured
and its operationalization.\looseness=-1

Using income as a proxy for SES would likely support some---though not
all---substantively interesting hypotheses involving SES. For example,
many social scientists have studied the relationship between SES and
health outcomes, demonstrating that people with lower SES tend to have
worse health outcomes. Measurements of SES obtained from the model
described in section~\ref{sec:income_as_proxy} would likely support
this hypothesis, albeit with some notable exceptions. For instance,
wealthy college students often have low incomes but good access
to healthcare. Combined with their young age, this means that they
typically have better health outcomes than other people with
comparable incomes. Examining these exceptions might reveal aspects of
the substantive nature of SES, such as wealth and education, that are
missing from the model described in
section~\ref{sec:income_as_proxy}.\looseness=-1

\subsubsection{Consequential validity}\label{subsubsec:consequential}

Consequential validity, the final aspect in our fairness-oriented
conceptualization of construct validity, is concerned with identifying
and evaluating the consequences of using the measurements obtained
from a measurement model, including any societal impacts.  Assessing
consequential validity often reveals fairness-related
harms. Consequential validity was first introduced by Messick, who
argued that the consequences of using the measurements obtained from a
measurement model are fundamental to establishing construct
validity~\cite{messick1987validity}. This is because the values that
are reflected in those consequences both derive from and contribute
back the theoretical understanding of the construct purported to be
measured.
In other words, the ``measurements both reflect structure in the
natural world, and impose structure upon
it,''~\cite{hand2016measurement}---i.e., the measurements shape the ways that we understand the construct itself.
Assessing consequential validity therefore means answering the
following questions: How is the world shaped by using the measurements?
What world do we wish to live in?
If there are contexts in which
the consequences of using the measurements would cause us to
compromise values that we wish to uphold,~then~the measurements should
not be used in those contexts.\looseness=-1

For example, when designing a kitchen, we might use measurements of a person's standing height to
determine the height at which to place their kitchen countertop. However, this may render~the
countertop inaccessible to them if they use a wheelchair.\looseness=-1

As another example, because the Universal Credit benefits system
described in section~\ref{sec:income_as_proxy} assumed that measuring
a person's monthly income by totaling the wages deposited into their
account over a single one-month period would yield error-free
measurements, many people---especially those with irregular pay
schedules---received substantially lower benefits than they were
entitled to.

The consequences of using scores obtained from value-added models are
well described in the literature on fairness in computational
systems. Many school districts have used such scores to make decisions
about resource distribution and even teachers' continued employment,
often without any way to contest these
decisions~\cite{amrein2008methodological,amrein2020methodological}.
In turn, this has caused schools to manipulate their scores and
encouraged teachers to ``teach to the test,'' instead of designing
more diverse and substantive curricula~\cite{oneil2016weapons}. As
well as the cases described above in sections~\ref{sec:test--retest}
and~\ref{sec:convergent}, in which teachers were fired on the basis of
low scores despite evidence suggesting that their scores might be
inaccurate, Amrein-Beardsley and Geiger
\cite{amrein2020methodological} found that EVAAS consistently gave
lower scores to teachers at schools with higher proportions of
non-white students, students receiving special education services,
lower-SES students, and English language learners. Although it is
possible that more effective teachers simply chose not to teach at
those schools, it is far more likely that these lower scores reflect
societal biases and structural inequalities. When scores obtained from
value-added models are used to make decisions about resource
distribution and teachers' continued employment, these biases and
inequalities are then exacerbated.\looseness=-1

The consequences of using scores obtained from COMPAS are also well
described in the literature on fairness in computational systems, most
notably by Angwin et al.~\cite{angwin2016machine}, who showed that
COMPAS incorrectly scored Black defendants as high risk more often
than white defendants, while incorrectly scoring white defendants as
low risk more often than Black defendants. By defining recidivism as
``a new misdemeanor or felony arrest within two years,'' COMPAS
fails to account for
false arrests or crimes that do not result in arrests. This assumption
therefore encodes and exacerbates racist policing practices, leading
to the racial disparities uncovered by Angwin et al.\ Indeed, by using
arrests as a proxy for crimes committed, COMPAS can only
exacerbate racist policing practices, rather than transcending
them~\cite{mayson2018bias,corbett2018mismeasure,green2020false,benjamin2019race,lum2016predict}. Furthermore,
the COMPAS documentation asserts that ``the COMPAS risk scales are
actuarial risk assessment instruments. Actuarial risk assessment is an
objective method of estimating the likelihood of reoffending. An
individual's level of risk is estimated based on known recidivism
rates of offenders with similar
characteristics''~\cite{recentCOMPASguide}. By describing COMPAS as an
``objective method,'' Northpointe misrepresents the measurement
modeling process, which necessarily involves making assumptions and is thus
never objective. Worse yet, the label of objectiveness obscures the
organizational, political, societal, and cultural values that~are
embedded in COMPAS and reflected in its consequences.\looseness=-1

Finally, we return to the high-risk care management models described
in section~\ref{subsec:patient-benefit}. By operationalizing greatest
care needs as greatest care costs, these models fail to account for
the fact that patients with comparable past care needs but different
access to care will likely have different past care costs. This
omission has the greatest impact on Black patients. Indeed, when
analyzing one such model, Obermeyer et al.\ found that only 18\% of
the patients identified for enrollment were
Black~\cite{obermeyer2019health}. In addition, Obermeyer et al.\ found
that Black and white patients with comparable future care costs did
not have comparable health statuses. In other words, these models
exacerbate the enormous racial health disparities in the U.S. as a
consequence of a seemingly innocuous assumption.\looseness=-1

\section{Fairness as a Construct}\label{sec:fairness}

We now explain how measurement modeling can contribute to recent
debates about fairness definitions. Although fairness feels
instinctively different to the constructs discussed so far, it is an
unobservable theoretical construct, albeit one about which there have
been millennia of disagreements. These disagreements reflect the fact
that fairness is an \emph{essentially contested
  construct}~\cite{gallie1955essentially,mulligan2016privacy}---i.e.,
fairness has multiple context-dependent, and sometimes even
conflicting, theoretical understandings. The contested nature of
fairness makes it inherently hard to measure: As we described in
section~\ref{subsubsec:content-validity}, some traditions make a
single theoretical understanding a prerequisite for establishing
content---and hence construct---validity. Other traditions (which we
draw on in our fairness-oriented conceptualization of construct
validity) abandon this prerequisite and instead simply require an
articulation of which understanding is being operationalized, arguing
that a failure to do this makes it difficult~to~meaningfully compare
different operationalizations.\looseness=-1

We start by discussing aspects of the substantive
nature of fairness that are necessarily missing from the quantitative,
parity-based operationalizations commonly found in the literature on
fairness in computational systems, focusing on the resulting threats
to both content and convergent validity. We then draw on well-known
examples from this literature to argue that although recent debates
about fairness definitions have been framed in terms of
operationalizations, they are, in fact, debates about different
theoretical understandings of fairness---i.e., debates about
values. We argue that by framing these debates in terms of
operationalizations, they are rendered less accessible to the
stakeholders that are most likely to be affected by the computational
systems in question. Finally, we touch briefly on the role of
demographic factors in measuring fairness.\looseness=-1

\subsection{The Substantive Nature of Fairness}
\label{subsec:justice-dueprocess}

To date, much of the literature on fairness in computational systems
has focused primarily on quantitative, parity-based
operationalizations of
fairness~\citep[e.g.,][]{dwork2012fairness,friedler2016possibility,hardt2016equality}. These
operationalizations are appealing to computer scientists because of
their quantitative nature and because they operate within the
boundaries of a single computational system, without reference to the
broader societal context in which the system is
situated~\cite{martin2020extending}. Yet these same properties mean
that these operationalizations necessarily lack aspects of the
substantive nature of fairness found in many of the theoretical
understandings long discussed by philosophers, lawyers, and social
scientists. For example, Arneson~\cite{arneson2018four} presents four
different, and sometimes conflicting, theoretical understandings of
equal opportunity---itself an understanding of fairness. Of these
understandings, it might be possible to operationalize at most one or
two using
quantitative, parity-based methods. Operationalizing the
others using such methods would lead to a lack of content validity.

We note that many theoretical understandings of fairness, including
some of the understandings discussed by
Arneson~\cite{arneson2018four}, depend on some notion of justice, such
as procedural justice, distributive justice, or representational
justice. Any operationalization of fairness that omits justice
entirely cannot therefore be thought of as wholly and fully capturing
the substantive nature of these understandings. Moreover, any such
operationalization would likely be ineffective at remedying historical
injustices---a threat to consequential validity.\looseness=-1

\vspace{-0.15cm}
\subsection{Individual Fairness vs. Group Fairness}
\label{subsec:indiv_group}

Putting aside the threats posed to content and consequential validity
by quantitative, parity-based operationalizations of fairness, we turn
next to the recent debates about fairness definitions found in the
literature on fairness in computational systems. In what was arguably
the first such debate, Dwork et al.~\cite{dwork2012fairness}
contrasted \emph{individual fairness}, which requires that similar people be
treated similarly, and \emph{group fairness}, which requires that different
groups of people, such as groups defined in terms of different
demographic factors, be treated similarly. Dwork et al.\ argued that
computational systems that satisfy some definition of group fairness
can still yield fairness-related harms for people belonging to those
groups. This argument has been discussed in many papers since
then~\citep[e.g.,][]{corbett2018mismeasure,joseph2016fairness,chouldechova2018frontiers},
forming an ongoing debate about individual fairness and group
fairness.\looseness=-1

Although this debate is usually framed in terms of mathematics---i.e.,
in terms of operationalizations---it is, at its core, a debate about
different theoretical understandings of fairness. Viewing the debate
through the lens of measurement modeling offers a new way to engage
with it. Individual fairness and group fairness are conflicting
theoretical understandings of fairness. In contrast to the former
understanding, the latter understanding fails to account for
fairness-related harms experienced by individual people. Any
operationalization of the latter understanding will necessarily lack
content validity if it is viewed as an operationalization of the
former. The debate is therefore about values---i.e., to what extent
should individual experiences matter?---not about the measurement
modeling process.\looseness=-1

\vspace{-0.15cm}

\subsection{Individual Fairness}

Even debates about different definitions of individual fairness are
often debates about different theoretical understandings. Individual
fairness requires that similar people be treated similarly. However,
the similarity of two people is itself an essentially contested
construct, with multiple context-dependent theoretical understandings.
Moreover, even when focusing on a single understanding, measuring the
similarity of two people is a nontrivial undertaking that involves
many different aspects of the human experience, including aspects that
are inherently subjective or personal. Any practical
operationalization of similarity will therefore likely lack
content---and possibly even
consequential---validity~\cite{hoffmannrawls}. Yet academic
papers about individual fairness often obscure or downplay this
reality.\looseness=-1

For example, although Dwork et al.\ acknowledge that similarity may
not be easy to measure, they argue that ``the [similarity] metric may
reflect the `best' available approximation as agreed upon by
society''~\cite{dwork2012fairness}. Despite giving an implicit nod to
the assumptions involved in the measurement modeling process,
including the assumption of which theoretical understanding to
operationalize, this definition obscures the difficulty of making
explicit and testing those assumptions by using the phrase ``as agreed
upon by society.''

As another example, Joseph et al.\ present a framework for
``ensuring'' individual fairness when using a computational system to
decide who should receive some resource or
opportunity~\cite{joseph2016fairness,joseph2016fair}. Crucially, their
framework assumes that recipients should be selected the basis of
their quality, which is further assumed to be observable and easy to
measure directly, yielding error-free measurements: ``[O]ur definition
of fairness...\ assumes the existence of an accurate mapping from
features to true quality for the task at hand''~\cite{joseph2016fair}. This assumption downplays the fact that quality
is an essentially contested construct, in turn downplaying the
difficulty of measuring the similarity of two people in terms of their
quality.

\vspace{-0.15cm}

\subsection{Group Fairness}\label{subsec:parity-fairness}

Like debates about individual fairness, debates about different
definitions of group fairness are often debates about different
theoretical understandings. The most well-known debate about group
fairness involves COMPAS, which operationalizes a defendant's risk of
recidivism using measurements of a variety of observable properties
(and other unobservable theoretical constructs), as we described in
section~\ref{subsec:recidivism}. The debate began when Angwin et
al.~\cite{angwin2016machine} showed that COMPAS incorrectly scored
Black defendants as high risk more often than white defendants, while
incorrectly scoring white defendants as low risk more often than Black
defendants. In other words, Angwin et al.\ showed that COMPAS was
unfair because it lacked \emph{error-rate
  balance}.\footnote{Error-rate balance is also known as
  \emph{equalized odds}~\cite{hardt2016equality}.} Northpointe
responded by arguing that COMPAS was fair because the probability of
recidivism among Black defendants scored as high risk was the same as
the probability of recidivism among white defendants scored as high
risk---i.e., COMPAS possessed \emph{predictive parity}. Prompted by
this, many academic papers began to discuss the properties and
shortcomings of different definitions of group fairness, with some
papers demonstrating that it is impossible for a system like COMPAS to
possess both error-rate balance and predictive parity when the
probability of recidivism for Black defendants is not the same as
the probability of recidivism for white
defendants~\citep[][]{kleinberg2016inherent,berk2017fairness,corbett2017algorithmic,corbett2018mismeasure,pleiss2017fairness,chouldechova2017fair}.

Despite being framed in terms of operationalizations, this debate is
actually about different theoretical understandings of
fairness. Indeed, in their discussion of the debate, Corbett et
al.\ wrote, ``At the heart of [the] disagreement is a subtle ethical
question: What does it mean for an algorithm to be
fair?''~\cite{corbett2016computer}. In other words, error-rate balance
and predictive parity are operationalizations of different theoretical
understandings of group fairness. These understandings---and hence
their operationalizations---have very different
consequences. Predictive parity means that a score has the same
meaning when it is given to a Black defendant as it does when it is
given to a white defendant. In contrast, error-rate balance means that
the errors experienced by Black defendants and white
defendants are comparable. Choosing one of these theoretical understandings over
the other therefore means choosing which values we wish to
uphold. But, by couching the debate in mathematics, this choice is
concealed~from~the stakeholders that are most likely to be affected by
it.\looseness=-1

\vspace{-0.15cm}

\subsection{Demographic Factors}\label{subsec:demographics}

Group fairness, as we explained in section~\ref{subsec:indiv_group},
requires that different groups of people, such as groups defined in
terms of different demographic factors, be treated similarly. But many
demographic factors, such as race or gender, are themselves
essentially contested constructs, with theoretical understandings that
vary across cultures and over time. As a result, measuring group
fairness often requires that we first undertake the non-trivial task
of measuring these constructs. Ideally, measurements of these
constructs should be self reported, thereby limiting the potential for
harmful errors. However, there are many contexts in which self-reports
at the level of individual people are not available or readily
accessible. In these contexts, it may be tempting to infer the
relevant constructs from measurements of observable properties (and
other unobservable theoretical constructs) thought to be related to
them---i.e., operationalize them via a measurement model. We emphasize
that even with careful consideration of different theoretical
understandings and careful assessment of both construct reliability
and construct validity, this approach is fraught and likely to cause a
variety of fairness-related and other
harms~\citep[e.g.,][]{goldenfein2018profiling,larson2017gender,olteanu2019social,keyes2018misgendering,bennett2020point},
despite good intentions. In other words, the consequences of using the
resulting measurements will~likely cause us to compromise values that
we wish to uphold.\looseness=-1

\section{Discussion}\label{sec:discussion}

Many computational systems involve unobservable theoretical
constructs---abstractions that describe phenomena of theoretical
interest, such as socioeconomic status, teacher effectiveness, and
risk of recidivism. Because these constructs cannot be measured
directly, they must instead be inferred from measurements of
observable properties (and other unobservable theoretical constructs)
thought to be related to them. However, this process---the measurement
modeling process---necessarily involves making assumptions, thereby
introducing the potential for mismatches between the theoretical
understanding of the construct purported to be measured and its
operationalization. As we argued in
section~\ref{subsec:construct-validity}, many of the harms studied in
the literature on fairness in computational systems are direct results
of such mismatches. This is because the assumptions made when moving
from abstractions to mathematics often encode~and exacerbate societal
biases and structural inequalities.\looseness=-1

Although measurement modeling is fundamental to the quantitative
social sciences, it has not traditionally played a role in computer
science. As a result, researchers and practitioners are often inclined
to collapse the distinctions between constructs and their
operationalizations, either colloquially or epistemically. But
collapsing these distinctions removes opportunities to anticipate and
mitigate fairness-related harms by eliding the space in which they are
most often introduced. Further compounding this issue is the fact that
measurements of unobservable theoretical constructs are often treated
as if they were obtained directly and without errors---i.e., a source of
ground truth. Measurements end up standing in for the constructs
purported to be measured, normalizing the assumptions made during the
measurement modeling process and embedding them throughout society. In
other words, ``measures are more than a creation of society, they
\emph{create} society''~\cite{alder2002measure}---a view also
expressed by Bowker and Star in their exploration of the consequences
of classification in computing systems~\cite{bowker2000sorting}.\footnote{Bowker and Star
  express the view that measurements create society via their
  discussion of classification in computational
  systems~\cite{bowker2000sorting}.  To see how classification in
  computational systems subtly but fundamentally creates categories
  and stratifications in the world, consider a user creating an
  account on a website. The website might require the user to select
  either ``Male'' or ``Female'' as their gender, refusing to create
  the user's account if one of these options is not selected. Bowker
  and Star argue that these kinds of design choices are fundamentally
  political: ``Seemingly purely technical issues like how to name
  things and how to store data in fact constitute much of human
  interaction and much of what we come to know as natural.'' As
  another example, many researchers have argued that race as a
  category further entrenches structural
  racism~\cite[e.g.,][]{benjamin2019race,hanna2020towards}.\looseness=-1}
Collapsing the distinctions between constructs and their
operationalizations is therefore not just theoretically or
pedantically concerning---it is practically concerning with very real,
fairness-related consequences. Moreover, because most computational
systems are developed by computer scientists, the practice~of
collapsing these distinctions is
widespread~\cite{passi2019problemformulation}.\looseness=-1

We argue that measurement modeling provides a both a language for
articulating the distinctions between constructs and their
operationalizations and set of tools---namely construct reliability
and construct validity---for surfacing possible mismatches. In
section~\ref{subsec:construct-validity}, we therefore proposed
fairness-oriented conceptualizations of construct reliability and
construct validity, uniting traditions from political science,
education, and psychology. We showed how these conceptualizations can
be used to 1) anticipate fairness-related harms that can be obscured
by focusing primarily on out-of-sample prediction, and 2) identify
potential causes of fairness-related harms in ways that reveal
concrete, actionable avenues for mitigating them. We acknowledge that
assessing construct reliability and construct validity can be
time-consuming. However, ignoring them means that we run the
risk of creating a world that we do not wish to live in.\looseness=-1

Finally, we turned to fairness itself, highlighting aspects of the
substantive nature of fairness that are missing from the quantitative,
parity-based operationalizations commonly found in the literature on
fairness in computational systems. We argue that although recent
debates about fairness definitions have been framed in terms of
operationalizations, they are, in fact, debates about different
theoretical understandings of fairness---i.e., debates about
values. Moreover, many definitions of fairness involve other
essentially contested constructs, such as similarity, quality, and
even demographic factors like race or gender. These relationships are
therefore reminiscent of the ``layers of bias'' described by Eckhouse
et al.~\cite{eckhouse2019layers} in the context of risk assessment
models: ``[E]ach layer depends on the layers below it: Without
assurances about the foundational layers, the fairness of the top
layers is irrelevant.'' Worse yet, by using such fairness definitions
to label computational systems as ``fair,'' we risk the adoption of
these systems without any critical assessment because their fairness
is assumed to be guaranteed.

To conclude, measurement modeling promotes greater transparency and
accountability by providing researchers and practitioners with a set
of tools for making explicit and testing assumptions. By using these
tools to surface mismatches between constructs and their
operationalizations, researchers will be better able to anticipate and
mitigate fairness-related harms arising from computational
systems. Although measurement modeling is largely missing from
computer science, we argue that it should be essential knowledge for
everyone developing or deploying computational systems.\looseness=-1

\section*{Acknowledgments}

We thank Solon Barocas, Su Lin Blodgett, Alex Chouldechova, Hal
Daum\'{e}, Lise Getoor, Moritz Hardt, Josh Kroll, Alexandra Olteanu,
Forough Poursabzi-Sangdeh, Brandon Stewart, Philip Thomas, Jenn
Wortman Vaughan, and many others for feedback on this
paper.\looseness=-1

\bibliographystyle{ACM-Reference-Format}
\bibliography{measurement}

\end{document}